# Automatic Multi-Stain Registration of Whole Slide Images in Histopathology

Abubakr Shafique[1], Morteza Babaie[1,3], Mahjabin Sajadi[1], Adrian Batten[2], Soma Skdar[2], and H.R. Tizhoosh[1,3]

*Abstract*— Joint analysis of multiple biomarker images and tissue morphology is important for disease diagnosis, treatment planning and drug development. It requires cross-staining comparison among Whole Slide Images (WSIs) of immuno-histochemical and hematoxylin and eosin (H&E) microscopic slides. However, automatic, and fast cross-staining alignment of enormous gigapixel WSIs at single-cell precision is challenging. In addition to morphological deformations introduced during slide preparation, there are large variations in cell appearance and tissue morphology across different staining. In this paper, we propose a two-step automatic feature-based cross-staining WSI alignment to assist localization of even tiny metastatic foci in the assessment of lymph node. Image pairs were aligned allowing for translation, rotation, and scaling. The registration was performed automatically by first detecting landmarks in both images, using the scale-invariant image transform (SIFT), followed by the fast sample consensus (FSC) protocol for finding point correspondences and finally aligned the images. The Registration results were evaluated using both visual and quantitative criteria using the Jaccard index. The average Jaccard similarity index of the results produced by the proposed system is 0.942 when compared with the manual registration.

## I. INTRODUCTION

Digital Pathology (DP) is becoming important tool for pathologists and paves the road for computational pathology [1]. In DP, multi-stain histopathological image analysis is critical for cancer diagnosis and prognosis [2]. Different cellular morphological maps and tissue expressions help the pathologists to analyse various functional or physical properties, which are important for cancer diagnosis [3], [4]. In prognosis of the disease, H&E stain slide is used as a gold standard to analyze tissue morphology; H causes the blue staining of nuclei whereas E causes the red/pink staining of cytoplasm and the extracellular connective tissue. On the other hand, immunohistochemistry (IHC) is used to detect antigens or proteins in biological tissues, allowing biologists and medical professionals to see exactly where a given protein is located within the tissue being examined [5]. IHC is an important tool in disease diagnosis, drug development and biological research as IHC staining is commonly adopted in the diagnosis of abnormal cells such as those found in cancerous tumours. With the knowledge of tumour markers and IHC tools, medical professionals are able to diagnose a cancer as benign or malignant, determine the stage and grade of a tumour and identify the cell type and origin of a metastasis in order to find the site of the primary tumour. Joint analysis of different IHC's are important to analyse any co-localised area, highlighting different useful information for the pathologists [6]. By registering the multiple images, we can see which biomarkers are co-expressed in the same part of the tissue. Currently, this registration process is manually done by the pathologists, which can be inaccurate, difficult, and time consuming [6]. In addition, registered mask with IHC could be used as a label for H&E images to apply AI technique such as image search [7] and network training [8] on the H&E images.

Image registration is the process of transforming different sets of data into one coordinate system, and elastic image registration is potentially an enabling technology for the effective and efficient use of many image guided diagnostic and treatment procedures, which rely on multi-modality image fusion or serial image comparison [9]. Alignment of cross-staining microscopic sections is challenging, especially when dealing with WSIs. WSI is a digital image which is scanned and recorded the tissue sample at high-resolution digitally [10]. At the highest resolution, WSI can exceed one billion pixels, and it is not feasible to conduct image alignment due to the limitation of computer memory. Apart from complex data deformations, such as morphological deformations, stain variations and imaging artifacts are introduced during slide preparation. Moreover, there are large variations on staining colors, cell appearance and tissue morphology across different slides. The aforementioned effects tend to make existing registration methods fail.

In previous studies, the BUnwarpJ approach [11] was developed for histological section alignment allowing image registration in bi-direction simultaneously. This approach is the extension of the Sorzano *et al.* [12], which was using b-splines for non-rigid registration. The LeastSquares method [13] was proposed for an image alignment technique with least square based registration method using linear feature correspondences. The Elastic approach [14] is image alignment technique based on non-linear block correspondences. There are other several 3D image registration techniques which allow alignment of one or more images on the serial biological images such as Wang *et al.* [9], Guyader *et al.* [15], Goubran *et al.* [16], and Tegunov *et al.* [17]. Guyader *et al.* [15] introduced groupwise registrations in single optimization procedure without reference image. Goubran *et al.* [16] presented Multi-modal Image Registration And Con-nectivity analysis (MIRACL) pipeline. Tegunov *et al.* [17] introduced Warp software that is developed for biological data acquisition at any cryo-EM facility and substantially speeds up the process of cryo-EM structure determination with improved results. However, these studies are not designed for the cross-staining alignment purposed.

[1]Abubakr Shafique, Morteza Babaie, Mahjabin Sajadi, and H.R. Tizhoosh are with Kimia Lab, University of Waterloo, Waterloo, ON, Canada. abubakr.shafique@uwaterloo.ca
[2]Adrian Batten and Soma Skdar are with Department of Pathology, Grand River Hospital, Kitchener, ON, Canada.
[3]Hamid Tizhoosh and Morteza Babaie are also affiliated with Vector Institute, MaRS Centre, Toronto, Canada.

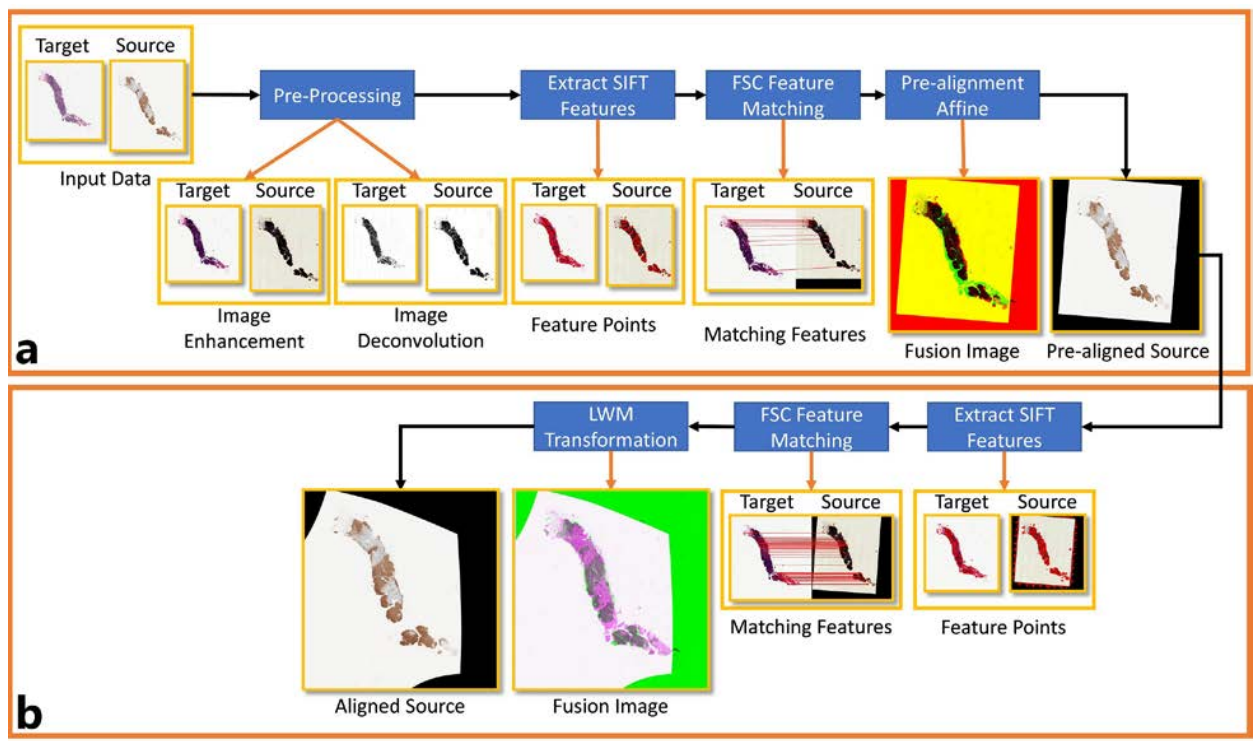

Fig. 1. Process flow of the proposed two step automatic registration method. (a) shows the first step of pre-alignment using affine transformation model. (b) shows the second step of automatic registration using local weight mean (LWM) for each control feature point.

Here, we present a two-step automatic cross-staining alignment of biological WSIs with different IHC stains and cross-staining alignment of H&E and IHC images for assisting medical experts to identify cancerous tissues and determine stage and grade of a tumour and perform multi-modal protein mapping. The proposed method is able to not only reduce global and local variations among slides but also effectively identify corresponding features and with accurately identified landmark features, the method produces valid registration outputs robust to various deformation problems commonly occurred in biological data, such as morphological distortions, staining variations, staining artifacts and loss of tissue. The rest of the paper is organized as follows: In Section II, the proposed method is presented. In Section III, the results from the proposed method are presented, and the description of the data used in this experiment. Section IV includes the discussion and the conclusion.

## II. MATERIALS & METHODS

In this paper, we proposed a two step automatic landmark based image registration method, which automatically detecting landmarks in both histopathological images (Source and Target), using the scale invariant image transform (SIFT) [18], [19] for feature detection and FSC [18] protocol for finding point correspondences. In the first step, source and target images are pre-aligned using a rigid affine transformation as shown in fig 1(a). In the second step, the pre-aligned images are registered together non-linearly using local weight mean (LWM) for each control feature point as shown in fig 1(b).

### A. Pre-alignment

In the first step, given a image pair, let $i_T(x, y)$ denote the target image, $i_S(x, y)$ denote the source image, and $i_P(\hat{x}, \hat{y})$ denote the registered source image. $i_S$ is rigidly transformed into $i_P$ using affine transformation model as

$$i_P(\hat{x}, \hat{y}) = A(i_S(x, y)), \quad (1)$$

where $i_P(\hat{x}, \hat{y})$ is the transformed (registered) source image, $A$ is the affine transformation model, and $i_S(x, y)$ is the source image which is to be transformed.

In order to compute affine transformation model, first, some pre-processing techniques are applied to enhance the contrast and features of our input images. Pre-processing includes image contrast enhancement and color deconvolution for both high level feature-based coarse registration and local area-based direct matching registration [9]. After pre-processing, key-point features are detected in a given image pair using SIFT [18]. SIFT is one of the most effective approaches to acquire the local descriptors from an image. After getting the key-points, usually RANSAC [20] method is used to remove the outlier descriptors, but in this approach we use a modified RANSAC approach called fast sample consensus (FSC) [18]. Here, FSC removes the outliers and keep the matching points based on the affine transformation aspect only. FSC is more accurate and fast than RANSAC and it can find correct matches from correspondence set in which even the majority of match pairs is false match. After removing the outliers and getting correct matching point pairs, affine transformation model is computed which is then applied to the source image ($i_S$).

*B. Non-linear Registration*

In the second step, pre-aligned source image ($i_P$) and original target image ($i_T$) are used as source and target image, respectively. The pre-aligned source image is aligned and non-rigidly registered to the target image using local weight mean (LWM) transformation model as

$$i_R(\bar{x}, \bar{y}) = L(i_P(\hat{x}, \hat{y})), \qquad (2)$$

where $i_R(\bar{x}, \bar{y})$ is the transformed (registered) pre-aligned source image, $L$ is the LWM transformation model, and $i_P(\hat{x}, \hat{y})$ is the pre-aligned source image from the step 1 which is to be transformed.

Pre-aligned source image is used for the non-rigid registration. Previously enhanced and color deconvolution images are transformed according to the pre-aligned transformation matrix, followed by the SIFT feature detection. SIFT extracts the enhanced features from the source ($i_P$) and target ($I_T$) images. In the feature extraction process, we divide the image into three portions and calculate the features and matching points in iteration using the same FSC method. Finally, we merge all the feature points which would give us feature points evenly distributed across the tissue. Then, LWM transformation model is computed based on the correct matching points. It is paramount for the LWM transformation model to have control points distributed all over the shape of the tissue, which results in an accurate transformation ($i_R$).

**Data –** For the experiment, 47 pairs of hodgkin lymphoma images are used including three IHC biomarkers (CD20, CD30, and PAX5), which are important for the Hodgkin's diagnosis and prognosis. The images were acquired from the Grand River Hospital, Kitchener, Ontario, Canada using Huron Tissue Scope LE scanner. The specimen are taken from different parts of the body. These specimens should be cut into small parts and each part will be fit to a paraffin cassette. From each cassette, H&E and all three IHCs are prepared on a glass slide, which are further scanned to obtain the digital WSI data. The experimental procedures involving human subjects described in this paper were approved by the Institutional Review Board.

## III. EXPERIMENTAL RESULTS

In order to evaluate the registration performance of the proposed method, we compared the automatic registration results with the manually registered results using Jaccard similarity index. Furthermore, automatically registered IHC results are also compared with the target H&E images using the the same Jaccard index.

For manual registration, corresponding points in the source and the target images were marked manually by the expert to calculate the LWM transformation function. Finally, the calculated LWM transformation function is then applied to the source image to get manually registered IHC image.

Jaccard similarity index for each pair is calculated to evaluate our results. Binary masks of the source images are transformed using the proposed automated LWM transformation function and manually computed LWM transformation function. After transforming the masks, the automated transformed source masks are compared with the manually transformed source masks to calculate the "Manual" jaccard index as shown in Fig. 2. Moreover, the automated transformed source masks are also compared with the corresponding H&E masks to calculate the "H&E" Jaccard index as shown in Fig. 2. Average similarity index of the automated and manually transformed IHC source images is 0.92, and average similarity index of the automated transformed IHC images and target H&E images is 0.94. In this experiment, three IHC biomarkers are aligned with their corresponding H&E stain. The three biomarkers include CD20, CD30, and PAX5, and the average jaccard similarity index of the automated registered biomarkers with manually registered biomarkers are 0.95, 0.93, and 0.94, respectively. On the other hand, the average Jaccard index of these three biomarkers when compared with their target H&E stain is 0.93 for CD20, 0.92 for CD30, and 0.93 for PAX5.

Additionally, the proposed method is also compared with two existing cross-stain registration methods (see Fig. 3). In our experiment, BUnwarpJ [11] was able to register 36 pairs, and Wang *et al.* [9] was only able to register 16 pairs out of 47. On the other hand, the proposed method was able to register all 47 pairs. A sample is shown in Fig. 3, where both methods failed to register the IHC but the proposed method is able to register successfully.

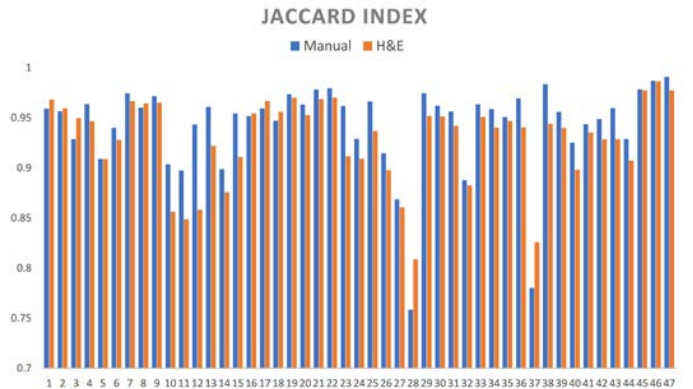

Fig. 2. The Jaccard similarity of each registered pair produced by the proposed automated system with same manually registered IHC image and target H&E image. Manual Jaccard index means the comparison of the automatically transformed source IHC image and manually transformed source IHC image. H&E Jaccard index means the comparison of the automatically transformed source IHC image and its corresponding H&E target image.

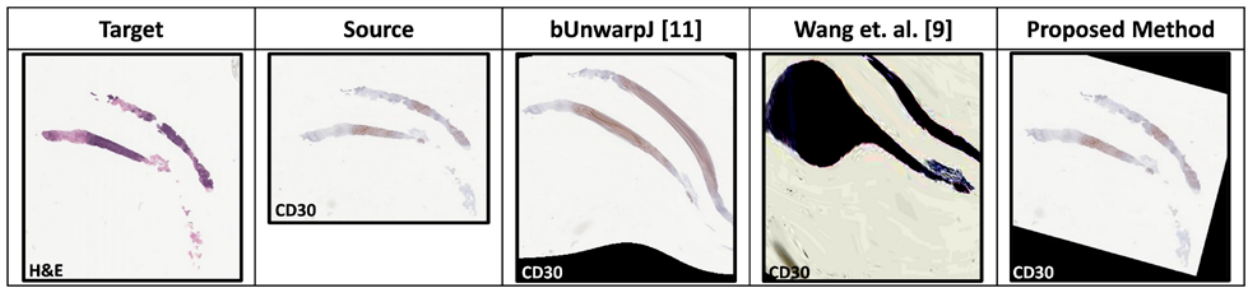

Fig. 3. A sample registration result is compared with the existing cross-stain registration methods.

## IV. Discussion & Conclusion

Multi-modal IHC mapping is important for disease diagnosis, drug development and biological research, requiring joint analysis of multiple protein expression maps and cellular morphology maps at single-cell resolution. Taking the diagnosis of Hodgkin Lymphoma (HL), in clinical practice multiple IHC images are used. For this reason, we proposed an improved registration method, which works even when the IHC images have very different morphology making other methods fails (see Fig. 3). Other methods tends to fail when there is difference in the morphology of the sections and the aspect ratio of the two images. The proposed method is not only useful for HL but also for the prostate, breast, and many other types of cancer which require multiple IHC image analysis. Automatic alignment of multi-staining tissue images, especially biomarkers with H&E image is helpful for screening lymph nodes sections and to identify potential metastatic tissues for further validation.

In conclusion, the proposed two-step method has shown promising results even on the complex data where other existing methods failed. For the future work, one could update the input pipeline to make the existing method adopt to 2D and 3D image registration.